\documentclass{appolb}
\usepackage{amsmath}
\usepackage[utf8]{inputenc}

\newcommand{\beqn}{\begin{equation}}
\newcommand{\eeqn}{\end{equation}}
\newcommand{\beq}{\begin{displaymath}}
\newcommand{\eeq}{\end{displaymath}}

\def\bal#1\eal{\begin{align}#1\end{align}}

\newcommand{\czf}[2]{f_{#1}\partial_{f_{#2}}}
\newcommand{\pad}[2]{\frac{\partial #1}{\partial #2}}
\newcommand{\bls}[1]{\boldsymbol{#1}}
\newcommand{\om}{\omega}
\newcommand{\vff}{\mathrm{F}_1^{-}}
\newcommand{\vfg}{\mathrm{F}_2^{-}}
\newcommand{\cD}{{\cal D}}

\allowdisplaybreaks
\begin{document}
 \eqsec  
\title{On collective octupole degrees of freedom -- \\
next pieces of the formal background
\thanks{Presented at XXVI Nuclear Physics Workshop "Marie \& Pierre Curie", Kazimierz Dolny, Poland
2019}%
}
\author{
{Leszek Pr\'ochniak}
\address{Heavy Ion Laboratory, University of Warsaw,\\ Pasteura 5A, PL-02-093 Warsaw, Poland}
}

\maketitle

\begin{abstract}

The concept of an intrinsic system can be extended to the case of collective
octupole degrees of freedom by exploiting the symmetry properties with respect to
transformations of the octahedral group $O_h$. Explicit formulas for scalar
invariants as polynomials of intrinsic variables are presented. A method of
constructing a basis in the space of functions on the octupole intrinsic
space is proposed.

\end{abstract}

{\it Dedicated to Krzysztof Pomorski on the occasion of the 50th anniversary of his
scientific activity.}


\section{Introduction}

The role played by collective octupole degrees of freedom still attracts
attention of both experimental and theoretical nuclear physicists. First of 
all,
one should mention attempts to find nuclei with a static octupole
deformation in actinides region~\cite{gaff}, however, as well lighter nuclei
($N=82$) are interesting subject of spectroscopic studies~\cite{komor}. On
the theoretical side, a proper description of such phenomena is difficult,
mainly due to the necessity of considering both quadrupole and octupole
deformations, see e.g. recent paper~\cite{dobrow}.

Here I present an extension of the results of \cite{acta,greiner} where we 
introduced the concept of an intrinsic system for the octupole space based on 
irreducible representations of the octahedral group $O_h$.

The paper is organized as follows. After a brief recollection of basic 
notions and facts about intrinsic octupole introduced in \cite{acta,greiner} 
I discuss rotational invariants built from these variables and show that, 
quite unexpectedly, the two variants of intrinsic variables are not fully 
equivalent. I then sketch a method of building  basis functions in the 
intrinsic frame which could in further perspective pave the way to a
construction of a basis in the full quadrupole plus octupole space.

\section{Intrinsic octupole coordinates}\label{octcoor}

Collective octupole degrees of freedom are described by variables which span
the seven-dimensional real irreducible representation of the rotation group
$SO(3)$ with the parity operation acting as multiplication by~$-1$. The best
known are seven complex variables $\alpha_\mu, \mu=-3,...3$ with the
additional condition $\alpha_{-\mu}=(-1)^{\mu}\alpha^*_{\mu}$ ensuring that
the considered irreducible representation of $SO(3)$ is of a real type. Another choice, also useful
in several applications, is to use seven real variables defined as
\beqn
a_0=\alpha_0, \ \ a_m=\sqrt{2}{\rm Re}\,\alpha_m, \ \ 
b_m=\sqrt{2}{\rm Im}\,\alpha_m, \  m=1,2,3 \ .
\eeqn

The concept of an intrinsic (of principal axes) frame of reference appeared
to be very fruitful in the case of quadrupole variables, however, extending of this
concept to the octupole case is not straightforward. To define an intrinsic
system for the octupole space we exploit properties of octupole variables
with respect to the octahedral group $O_h \subset O(3)$, for more details
see~\cite{acta,greiner}. This group leaves invariant a set of three
intersecting perpendicular lines (without fixed directions and labels) which
seems to be a fundamental property of a reference frame. The octupole space
can be decomposed as $\mathrm{A}_2^{-}\oplus \vff\oplus \vfg$ of irreducible
representations of $O_h$, with dimensions $1,3,3$, respectively,
see~\cite{ham}. Coordinates compatible with this decomposition are denoted
as $(b,\bls{f},\bls{g})=(b,f_x,f_y,f_z,g_x,g_y,g_z)$, where $b$, $f_k$,
$g_k$, $k=x,y,z$ span representations ${A}_2^{-}$, $\vff$, $\vfg$,
respectively. Explicit expressions for them are given e.g. in Appendix~B
of~\cite{greiner}.

In \cite{acta,greiner} we introduced two variants of the intrinsic system:
$\vff$-type and $\vfg$-type. In the case of $\vff$ variant instead of the
LAB system $(b,\bls{f},\bls{g})$ variables one uses
$(b',\bls{f'},\bls{\omega})$ variables where $\bls{\omega}$ are the Euler
angles of the rotation such that the result of this rotation has the form 
$(b',\bls{f'},0)$.
The $\vfg$ case is defined analogously, with $\bls{f}$ and $\bls{g}$
interchanged. We presented detailed
consequences of this change of variables, in particular we discussed the
Hamiltonian and the angular momentum operator expressed through intrinsic
variables. It appears, however, a bit surprisingly, that the $\vff$ and
$\vfg$ variants are not equivalent, in particular the $\vfg$ variant does
not cover the whole octupole space, as will be shown in
Section~\ref{sect:rotinvar}. In the following I drop primes and denote
intrinsic coordinates as $(b,\bls{f},\bls{\om})$. The $(b,\bls{f})$,
$\bls{\om}$ will be
called the deformation and rotation part, respectively, of octupole
variables.

\section{Rotational invariants}\label{sect:rotinvar}

Of particular importance 
in the analysis of functions of collective variables are 
the simplest building blocks of a given multipolarity, in 
other words the simplest polynomials (of a given multipolarity) built from
the variables. For example, in the quadrupole case all scalar polynomials
can be built from two well known invariants $\beta^2$, $\beta^3\cos 3
\gamma$. The octupole case is more difficult. Exploiting the connection
between invariants of binary forms, see e.g.~\cite{olver} and scalar
polynomials of multipole tensors one finds that in the octupole case there
are 4 elementary scalars of the order 2, 4, 6 and 10, which generate
 the full ring of polynomial scalars. Moreover, there is also a
pseudoscalar polynomial of the order 15. For details see~\cite{oscoct},
where the laboratory frame was discussed. The scalars, denoted here by
$\eta_{2,4,6,10}$, can be expressed through couplings of the $\alpha$
variables as follows:
\bal
&\eta_2=-\sqrt{7}[\alpha \alpha]^0 \ ,\\
&\eta_4=\alpha^{4[2,3,0]} \ ,\label{eq:alf4}\\
&\eta_6=\alpha^{6[2,1,2,3,0]}, \label{eq:alf6}\\
&\eta_{10}=\alpha^{10\,[2,1,2,1,2,1,2,3,0]} \ , \label{eq:alf10}
\eal
where $\alpha^{4[2,3,0]}= 
[[[\alpha\alpha]^2\alpha]^3\alpha]^0$. The analogous notation is used in 
(\ref{eq:alf6},\ref{eq:alf10}).
Below we show 
explicit formulas for elementary scalars for both $\vff$ and $\vfg$ variants of the 
intrinsic frame.

\medskip
\noindent $\vff$ {\bfseries\itshape variant}.
\bal
&\eta_2=b^2+f^2_x+f^2_y+f^2_z=b^2+\sigma_2\label{eq:etaf2}\ , \\
&\eta_4=
\displaystyle
\frac{1}{84\sqrt{5}}\big(16\sigma_4-13\sigma_{42}+
80b^2\sigma_2 +24\sqrt{15}b\sigma_3\big)\label{eq:etaf4}\ , \\
&\eta_6=
\frac{\sqrt{3}}{196}\left((11\sigma_2\sigma_{42}-15\sigma_3^2)-\frac{16}{5}\sigma_2^3\right)
-\frac{1}{7\sqrt{5}}b\sigma_3\sigma_2+\\
&+\frac{1}{196\sqrt{3}}b^2(32\sigma_2^2-129\sigma_{42})
+\frac{2\sqrt{5}}{7}b^3\sigma_3
-\frac{20}{147\sqrt{3}}b^4\sigma_2\label{eq:etaf6}\ ,
\eal
\bal
& 9604\, \eta_{10}=
-\frac{400}{9\sqrt{3}}b^8\sigma_2+
\frac{184\sqrt{5^3}}{3}b^7\sigma_3
-\frac{5}{\sqrt{3^3}}b^6 (64\sigma_{2}^2+673\sigma_{42})+ \nonumber\\
&+\frac{2092\sqrt{5}}{3}b^5 \sigma_{3}\sigma_{2} 
+\frac{1}{\sqrt{3}}b^4(32\sigma_{2}^3-715\sigma_{42}\sigma_{2}-2289\sigma_{3}^2)+\nonumber\\
&+\frac{16}{\sqrt{5}}b^3\sigma_{3}(5\sigma_{2}^2+67\sigma_{42}) 
+\frac{\sqrt{3}}{5}b^2(192\sigma_{2}^4-439\sigma_{42}\sigma_{2}^2-691\sigma_{3}^2\sigma_{2}-270\sigma_{82})+\nonumber\\
&-\frac{3}{\sqrt{5}}b 
\sigma_{3}(84\sigma_{2}^3-259\sigma_{42}\sigma_{2}+255\sigma_{3}^2)+\nonumber\\
&-\frac{3\sqrt{3}}{100}(576\sigma_{2}^5-2700\sigma_{42}\sigma_{2}^3+2025\sigma_{82}\sigma_{2}+7350\sigma_{3}^2\sigma_{2}^2-1625\sigma_{3}^2\sigma_{42}) \ ,
\eal
where auxiliary symmetric functions of $f_x,f_y,f_z$ are defined as
\bal
\sigma_2&=f_x^2+f_y^2+f_z^2
&\sigma_3&=f_xf_yf_z\ , \\
\sigma_4&=f^4_x+f^4_y+f^4_z
&\sigma_{42}&=f^2_x f_y^2+f^2_yf^2_z+f^2_xf^2_z\ , \\
\sigma_6&=f^6_x+f^6_y+f^6_z
&\sigma_{82}&=f^4_x f_y^4+f^4_yf^4_z+f^4_xf^4_z\  .
\eal

\medskip
\noindent $\vfg$ {\bfseries\itshape variant.}

\bal
&\eta_2=b^2+g^2_x+g^2_y+g^2_z=b^2+\tau_2\label{eq:etag2}\ ,\\
&\eta_4 =  
\displaystyle
\frac{5\sqrt{5}}{28}\tau_{42}\label{eq:etag4} \ ,\\
&\eta_6=
\displaystyle
\frac{25\sqrt{3}}{196}\big(3\tau_3^2-\frac{1}{3}(b^2+\tau_2)\tau_{42}\big)\label{eq:etag6}
\ ,
\eal
\bal
&9604\, \eta_{10}=
-\frac{125}{3^{3/2}}\bigg[b^6\tau_{42}+3b^4(\tau_2\tau_{42}-9\tau_3^2)+
3b^2(\tau_2^2\tau_{42}+2\tau_{42}^2-15\tau_3^2\tau_2)+\nonumber \\
&-9b\tau_{63}\tau_3 
 +\big(\tau_2^3\tau_{42}+\frac{33}{4}\tau_{42}^2\tau_2-9\tau_3^2(4\tau_{2}^2-\frac{3}{4}\tau_{42})\big)\bigg]
\ ,
\eal
where symmetric functions of $g_x,g_y,g_z$ are defined as:
\bal
\tau_2&=g_x^2+g_y^2+g_z^2 &\tau_3&=g_xg_yg_z \ ,\\
\tau_{42}& =g^2_x g_y^2+g^2_yg^2_z+g^2_xg^2_z& 
\tau_{63}&=(g^2_z- g_x^2)(g^2_z-g^2_y)(g^2_y-g^2_x)\ .
\eal

One should remember that the generators of the ring of invariants (in the
polynomial sense) are not uniquely defined. First, they can be multiplied by
real numbers. Second, e.g. by adding the product $\eta_2\eta_4$ to $\eta_6$ we
obtain again a sixth order scalar (with respect to rotations).

As a simple exercise in the application of invariants one can check that a point
which in the LAB system has coordinates $(b=0,0,0,f_z\ne0,0,0,0)$ and is
trivially described in the $\vff$ intrinsic system does not belong to the
space of the $\vfg$ system. The values of invariants does not depend on the
choice of the coordinate system and by applying the formulas
(\ref{eq:etaf2}-\ref{eq:etaf6}) and
(\ref{eq:etag2}-\ref{eq:etag6}) we obtain
an evident contradiction  $\tau_{3}^2=-\frac{64}{3375}f_z^6$. Similar
considerations show that the set of points of the octupole space which are
not covered by the $\vfg$ is even larger than discussed in this paragraph.

\section{Basis in the intrinsic space}\label{sect:baseint}

In order to apply the presented formalism in the nuclear theory, e.g. for
study of Hamiltonian and other operators, one requires one more important
component, namely an appropriate basis in the Hilbert space of functions
defined on the octupole space. Again, the construction of such a basis is more
difficult than in the quadrupole case, where several approaches were
successfully applied~\cite{Pro09}. As can be seen in~\cite{oscoct} building
eigenfunctions of the harmonic oscillator which have good angular momentum
numbers is very difficult even in the laboratory system. Here I follow the
general idea of the method applied in~\cite{Pro99} for the quadrupole case
and which can be summarized as follows. From a properly chosen dense (in the
sense of the Hilbert space) set of functions of the intrinsic octupole
variables one constructs a subspace of functions which fulfill two
conditions. First (A), they are invariant with respect to the action of the octahedral
group. Second (B), they belong to the domain of the Laplace-Beltrami
operator which is compatible with the scalar product induced in the
intrinsic system by the standard Cartesian scalar product in the laboratory
system. From the physical point of view this operator is (up to a constant
factor) the simplest form of the kinetic energy operator in the intrinsic
system, see~\cite{acta,greiner}. Then, the standard orthonormalization
methods can be applied. At present the above procedure has been applied only
to the $\vff$ variant, with results presented briefly below.

In the deformation part of the octupole space we take all polynomials of the
$(b,\bls{f})$ variables times the Gaussian factor $\exp(-(b^2{+}\bls{f}^2)/2)$.
Considering a more general exponent $-c^2(b^2{+}\bls{f}^2)/2$
does not present any difficulty. 

{\bfseries\itshape Condition A.} Symmetrization.

It turns out that in the symmetrization stage it is more convenient to use
the spherical coordinates for $\bls{f}$, which are denoted by
$t,\xi_1,\xi_2$. Then the basic set of functions is organized as follows:
\bal
& \label{eq:btyd}b^{n_b}t^{n_t}Y_{lm}(\xi_1,\xi_2)e^{-(b^2+t^2)/2}\,\cD^{J}_{MK}(\boldsymbol{\om})
\ ,\\
& N=n_b+n_t, \ N=0,1,2, ..., \ \  l=n_t, n_t-2,...,0 (1) \nonumber \ ,
\eal
where $\cD^{J}_{MK}$ are the rotation matrices providing the basis in the
$L^2(SO(3))$ space.
One should keep in mind that $(lm)$ indices refer to the group
$SO_{\Lambda}(3)$ generated by the  operators (\ref{eq:lambda}) and not the
"main" $SO(3)$ mentioned in Section~2. However, thanks to vector-type
properties of the $\vff$ IR of the group $O_h$ the action of this group on
the functions $Y_{lm}(\xi_1,\xi_2)$ can be easily determined. Let us add
that $t$ and the Gaussian factor are invariant against $O_h$ and $b$ is a
pseudoscalar with respect to the $O_h$ action. To obtain $O_h$-invariant functions
from (\ref{eq:btyd}) we apply the projection operator
\beqn
P_3P_2P_1=(I+R_3+R_3^2)(I+R_2+R_2^2+R_2^3)(I+R_1) \ , 
\eeqn
where $R_k, k=1,2,3$ are well known generators of $O_h$, see
e.g.~\cite{margetan}. If we use $btY\cD$ as a shorter notation for
(\ref{eq:btyd}), with the Gaussian factor skipped, we arrive at the formulas
\bal
P_1\ btY\cD&=bt\ (Y_{lm}\cD^J_{MK}+ (-1)^{l+J}Y_{l-m}\cD^J_{M-K}) \ ,\\
P_2\ btY\cD&=[1+(-1)^{m+K}][1+(-1)^{n_b+(m+K)/2)}]\ btY\cD \ , \\
P_3\ btY\cD&=bt
\textstyle\sum_{m',K'}\big[\delta_{mm'}\delta_{KK'}+\big(i^{m{+}K}{+}(-i)^{m'{+}K'})d^l_{m'm}(\textstyle\frac{\pi}{2})
d^J_{K'K}(\textstyle\frac{\pi}{2})\big]\times \nonumber\\
&\quad\quad \times Y_{lm'}\cD^J_{MK'} \ ,
\eal
where $d^l_{mn}(\theta)$ is the "small" Wigner function. Of course, after 
the 
projection one should choose linearly independent functions, what is rather
easy in the considered case. In particular, one can obtain a simple
analytical formula for a number of such functions for given $(n_b,n_t,l,J)$.

{\bfseries\itshape Condition B.} Domain of the Laplace-Beltrami operator. 

At this stage we turn back from spherical to Cartesian coordinates for
$\bls{f}$. The Laplace-Beltrami operator can be written as (using slightly
more compact notation than in~\cite{acta,greiner})
\beqn
\Delta=
\frac{1}{d}\pad{}{b}d\pad{}{b}+
\sum_{s=x,y,z} \frac{1}{d}\pad{}{f_s}d\pad{}{f_s}+
\frac{1}{d}\sum_{k,j=1,2,3}W_{k}dM_{kj}W_{j} \ ,
\eeqn
where 
\beqn
d=8\left(b^3-(15/16) b(f_{x}^2+f_y^2+f_{z}^2)+2(15/16)^{3/2} f_{x} f_y f_z\right)
\eeqn
is the deformation part of the Jacobian of the change of variables (from the laboratory system to
the intrinsic system, \cite{greiner}, eq.~(13)) and
\bal
&M=4{\footnotesize
\begin{pmatrix}
{{16b^2+15({f_{y}}^2{+}
 {f_{z}}^2)}}
&{{15{f_{x}}{f_{y}}+8\sqrt{15}b
 {f_{z}}}}&{{8\sqrt{15}b{f_{y}}+15{f_{x}}
 {f_{z}}}}\cr 
{{15{f_{x}}{f_{y}}+8\sqrt{15}
 b{f_{z}}}}&{{16b^2+15({f_{x}}^2{+}{f_{z}}^2)
 }}&{{8\sqrt{15}b{f_{x}}+15{f_{y}}{f_{z}}}}\cr 
{{8\sqrt{15}b{f_{y}}+15{f_{x}}{f_{z}}
 }}&{{8\sqrt{15}b{f_{x}}+15{f_{y}}{f_{z}}}}&{{16b^2+15({f_{x}}^2{+}{f_{y}}^2)}}
\end{pmatrix}^{-1}
} \ ,\\
&W_{k}=S_k(\boldsymbol{\om},\partial_{\boldsymbol{\om}})+
\Lambda_{k}(\boldsymbol{f},\partial_{\boldsymbol{f}}), \ k=1,2,3 \ ,
\eal
with 
\beqn\label{eq:lambda}
\Lambda_{1}=3(\czf{y}{z}-\czf{z}{y})/2, \ \ \mbox{etc.}
\eeqn
and with $S_k$ being components of the angular momentum multiplied by $i$ (see
\cite{greiner}, eq.~(16)).

It can be verified that the functions obtained through symmetrization in the
previous stage, denoted as 
\beq
G(b,\bls{f},\bls{\omega}) e^{-(t^2+b^2)/2} \ ,
\eeq
belong to the domain of $\Delta$  provided that
\beqn
d^2\,\sum_{\beta=b,\bls{f}}\partial_\beta (d\partial_\beta G)
+d\,\sum_{kj}W_k ( d^2M_{kj} W_j G)
-\sum_{kj} (W_k d) d^2M_{kj}W_jG
\eeqn
can be expressed as
\beqn
d^3 p(b,\bls{f}) \ ,
\eeqn
where $p$ is a polynomial (0 non excluded). This condition is
far from trivial, even for low values of $N$ (the order of $G$ in the
$(b,\bls{f})$ variables), say $N=4$, we have to deal with polynomials of
order greater than 10. Hence, to find linear combination of the symmetrized
functions that fulfill condition (B) I have used the Maxima system for
symbolic computations. The developed procedures can be applied for $N\le
10$, $J\le 12$ but at present the more detailed analysis has been done for $N$ up
to 4 and $J$ up to 6.

A few remarks and examples. 

1. $J{=}0$. For even $N$ the conditions (A) and (B) are fulfilled by
polynomials $\eta_2^{N/2}=(b^2{+}\bls{f}^2)^{N/2}$ from which one can easily
build Laguerre polynomials $L^{(5/2)}_{N/2}$, which enter the eigenfunctions
of the octupole harmonic oscillator with the seniority number $\lambda=0$,
see~\cite{oscoct}.

2. $N{=}4,J{=}0$, in this case we have 5 functions after stage (A)
from which 2 linear combinations fulfilling condition (B) can be built. One
is, as expected, proportional to $\eta_2^2$ while the second one is
proportional to $\eta_4$, eq.~(\ref{eq:etaf4}). This is again a 
nontrivial fact, because this result is obtained in a way that is completely
independent from the theory applied in Section~3.

3. The rotation part of the basis functions is conveniently expressed using
the semi-Cartesian Wigner functions, see \cite{greiner}, eq.~(8):
\bal
&D^{J(+)}_{MK}=\big({\cal D}^{J}_{MK}+(-1)^K {\cal
D}^J_{M,-K}\big)/\sqrt{2(1{+}\delta_{K0})},  \ \ K\ge 0 \ ,\\
&D^{J(-)}_{MK}=i\big({\cal D}^{J}_{MK}-(-1)^K {\cal D}^J_{M,-K}\big)/\sqrt{2}, \ \ K>0 \
.
\eal
For example, in the case $N=2,J=2$ there is only one basis function which can be
written as
\beqn
\Psi^2_{2M}=\sum_{K=0}^{2}C^{(+)}_K D^{2(+)}_{MK}+\sum_{K=1}^{2}C^{(-)}_K D^{2(-)}_{MK}
 \ ,
\eeqn
with
\bal
C^{(+)}&=
\left\{ 2f_{z}^2{-}f_{y}^2{-}f_{x}^2 ,\, (\sqrt{3}
 f_{x}f_{z}{+}4\sqrt{5}bf_{y})/2 ,\, -\sqrt{3}\left(f_{y}{-}
 f_{x}\right)\left(f_{y}{+}f_{x}\right) \right\}
 \ , \\
C^{(-)}&=\left\{ \sqrt{3}f_{y}f_{z}{+}4\sqrt{5}bf_{x}
  ,\,\, -4\sqrt{5}bf_{z}{-}\sqrt{3}f_{x}f_{y} \right\}.
\eal

\section{Concluding remarks}

The results presented in this contribution extend our knowledge on the
formal properties of the intrinsic system for the octupole tensors. However,
there are still problems which remain only partly solved. Let us mention 
two.
First, despite some hints there is no formal proof that the variables
chosen according to the $\vff$ variant cover the whole octupole space.
Second, the integration measure in the deformation part contains the factor
$|d|=|8\left(b^3-(15/16) b(f_{x}^2+f_y^2+f_{z}^2)+2(15/16)^{3/2} f_{x} f_y
f_z\right)|$ which seems to make it impossible to obtain analytical results
for a scalar product of even the simplest basis functions and requires a very careful
numerical treatment. On the other hand, limitations on $N$ and $J$
for the basis discussed in Section~4 seem to be not very important for
physical applications and, moreover, with some more work, the upper limits
can be raised.

Inspiring discussions with S.G.~Rohoziński during our long-lasting collaboration are
gratefully acknowledged.

\end{document}